\documentstyle[11pt,aaspp4,tighten]{article}

\slugcomment{CNU-AST-96}

\begin{document}

\font\twelvei = cmmi10 scaled\magstep1 
       \font\teni = cmmi10 \font\seveni = cmmi7
\font\mbf = cmmib10 scaled\magstep1
       \font\mbfs = cmmib10 \font\mbfss = cmmib10 scaled 833
\font\msybf = cmbsy10 scaled\magstep1
       \font\msybfs = cmbsy10 \font\msybfss = cmbsy10 scaled 833
\textfont1 = \twelvei
       \scriptfont1 = \twelvei \scriptscriptfont1 = \teni
       \def\mit{\fam1 }
\textfont9 = \mbf
       \scriptfont9 = \mbfs \scriptscriptfont9 = \mbfss
       \def\bmit{\fam9 }
\textfont10 = \msybf
       \scriptfont10 = \msybfs \scriptscriptfont10 = \msybfss
       \def\bmsy{\fam10 }

\def\etal{{\it et al.~}}
\def\eg{{\it e.g.}}
\def\ie{{\it i.e.}}
\def\lsim{\raise0.3ex\hbox{$<$}\kern-0.75em{\lower0.65ex\hbox{$\sim$}}} 
\def\gsim{\raise0.3ex\hbox{$>$}\kern-0.75em{\lower0.65ex\hbox{$\sim$}}} 
 
\title{Numerical Simulations of Standing Shocks in Accretion Flows\\
   around Black Holes: A Comparative Study\footnote[1]
   {The Astrophysical Journal in press}}
 
\author{Diego Molteni}
\affil{Istituto di Fisica , Via Archirafi 36, 90123 Palermo Italy;\\
   e-mail: molteni@gifco.fisica.unipa.it}
\author{Dongsu Ryu}                            
\affil{Dept. of Astronomy \& Space Sci., Chungnam National Univ., 
   Daejeon 305-764, Korea;\\
   e-mail: ryu@sirius.chungnam.ac.kr}
\and
\author{Sandip K. Chakrabarti}
\affil{Tata Institute of Fundamental Research, Bombay 400005, India;\\
   e-mail: chakraba@tifrc2.tifr.res.in}

\vskip 1cm
\begin{abstract}

We compare the results of numerical simulations
of thin and quasi-spherical (thick) accretion flows 
with existing analytical solutions. We use a Lagrangian 
code based on the Smooth Particle Hydrodynamics (SPH) scheme and an Eulerian 
finite difference code based on the Total Variation Diminishing (TVD) scheme.
In one-dimensional thin flows, the results of the simulations,  with or without
shocks, agree very well with each other and with analytical solutions.
In two-dimensional thick flows, the general features, namely the 
locations and strengths of centrifugal and turbulent pressure 
supported shocks, centrifugal barriers, and the funnel walls which are
expected from analytical models, agree very well, though the 
details vary. Generally speaking, the locations of the shocks may be better
obtained by SPH since the angular momentum is strictly preserved in SPH,
but the shocks themselves are better resolved by TVD.
The agreement of these code test results with 
analytical solutions provides us with confidence to apply these codes
to more complex problems which we will discuss elsewhere.
 
\end{abstract}


\clearpage
 
\section{INTRODUCTION}
 
A great deal of interest is present in the astrophysical 
community in understanding
the accretion flows around black holes, both analytically as well
as numerically. The scope of analytical works is usually very limited, 
since these are devoted to obtain only steady state 
solutions or to study stability properties
of these solutions based on local or in rare cases, global
stability analysis. However, astrophysical systems
are hardly in a steady state. Fluctuations and flickering 
of radiation emitted from these
systems constantly remind us of time-variations of the 
dynamical and thermo-dynamical 
quantities. The variations can take place on times scales of 
few micro-seconds to few years.
In order to understand such systems it is 
necessary to perform numerical simulations of these systems. 
On the other hand, numerical techniques of 
writing a code are not unique, and 
not all techniques are equally accurate. 
To be able to rely on a particular numerical
scheme, it is required that numerical results be tested against analytical
solutions, stationary or non-stationary, 
of presumably simpler systems which are as similar as possible
to the actual systems that will be eventually studied by the code. Testing of
a code is thus an essential component of a numerical study 
of the long time behavior of
a system. In the present paper, we shall compare two codes, one is
a Lagrangian code based on the Smoothed Particle Hydrodynamics 
(SPH; Monaghan, 1985, 1992) 
and the other is an Eulerian code based 
on the Total Variation Diminishing scheme (TVD;
Harten 1983; Ryu \etal 1993, 1995) appropriately modified 
to study cylindrically symmetric problems.
The tests are carried out against some well known non-linear
solutions of axisymmetric transonic accretion disks around
a black hole. We show that both
of our codes agree very well with the analytical results. Armed with the
confidence that our codes are reasonably good 
for study of time dependent flows,
we present more complex behavior of time dependent 
accretion flows in our next papers
(Ryu, Chakrabarti, \& Molteni 1996; 
Molteni, Chakrabarti, \& Ryu 1996). 

Accretion flows are important ingredients in many astrophysical systems
which involve mass transfer from one object to another (such as
in a binary system) or set of objects to another 
(such as in a galactic center).
There is ample evidence in the literature that accretion disks exist
in systems ranging from CVs, LMXBs on a small scale, to AGNs and Quasars
on a large scale. Standard disk models 
of Shakura \& Sunyaev (1973) and Novikov \& Thorne
(1973) assume Keplerian distribution of accreting matter. 
Here, the inner edge of the
disk is chosen to coincide with the marginally stable orbit located at three
Schwarzschild radii ($r_i=3r_g$, where $r_g=2GM_{BH}/c^2$ 
is the Schwarzschild radius of a black hole of mass $M_{BH}$). 
These disk models are clearly incomplete, since
the inner boundary condition on the horizon was not taken care of. 
Because flow at the inner boundary (namely, the horizon)
is bound to be supersonic, in order to satisfy this condition, it
would start to deviate from a Keplerian disk and enter into the
black hole horizon through the sonic surface or sonic horizon even if it 
originally starts from a Keplerian disk. 
Paczy\'nski \& Bisnovatyi-Kogan (1981) generalized the equations governing these
flows and solved these equations 
only close to the inner edge of the disk using numerical methods. 
It was demonstrated that indeed the flow enters the
black hole supersonically. 

This is not the full story. Liang \& Thompson (1980) pointed out that unlike
a Bondi flow, where only one saddle type sonic point is allowed
(e.g., Thompson \& Stewart, 1985), rotating accretion and 
winds can have two saddle type
sonic points. After the flow passes through the outer sonic point,
the centrifugal force can virtually stop the rotating supersonic
flow after forming a shock close to a black hole 
and restart its journey once more to pass through the inner sonic point
and to satisfy the inner boundary condition on the horizon (Chakrabarti 1989, 
hereafter C89). Extensive analytical results
are now available to understand the nature of these transonic solutions
which contain standing shocks (C89) and
recently, it has been shown that these generalized accretion flows could be 
responsible for the hard and soft state transitions 
in the black hole candidates (Chakrabarti \& Titarchuk, 1995), 
or even responsible for quasi-periodic oscillation (QPO)
of the hard X-rays from the black hole candidates (Molteni, Sponholz \& 
Chakrabarti 1996, hereafter MSC96; Ryu \etal 1995). In latter work,
one invoked resonance oscillation of the shocks to explain QPOs. Recently,
we found that the oscillation is more generic in that it is related to
the transient shock formation in a wider parameter space (Ryu, Chakrabarti, Molteni, 
1996) thus suggesting persistent QPOs even when accretion rate varies by
a large factor.

Numerical simulations of the shocks in accretion flows, however,
preceded these analytical studies. The first numerical attempt 
to study the behavior of rotating accreting matter around black holes
was made by Wilson (1978). An Eulerian, fully general relativistic,
first-order backward space difference technique was used.
The spatial resolution was low and the solution was evolved only 
upto $\sim 100 GM/c^3$. It was shown that large 
angular momentum accretion is 
accompanied by shocks which traveled outwards. This code was later improved
upon, with the number of grid points as well as the evolution time 
orders of magnitude higher. A series of very important simulations were made
with this code to show that thick accretion disks can indeed form in inviscid 
flows (Hawley, Smarr \& Wilson 1984, 1985). 
These simulations also confirm the results of Wilson (1978)
that non-steady shocks are formed which travel outward. From 
the post-shock flow, a very strong wind is generated which is hollow
and hugs the funnel wall. Due to the inviscid nature of the simulation,
the centrifugal force kept the flow away from the axis of symmetry.
A better understanding of the shock formation 
in the disk emerged only after the
analytical works and numerical works (C89, Chakrabarti \& Molteni 1993, 
hereafter CM93; Molteni, Lanzafame \& Chakrabarti 1994, 
hereafter MLC94) were able to show that standing 
shocks can not only form in accretion and winds, they are stable as well.
They also form very close to the location where analytical works predict them.
Subsequently, these numerical simulations were continued in the case of flows
around a Kerr black hole (Sponholz \& Molteni, 1994, hereafter SM94) 
and in viscous transonic disks (Chakrabarti \& Molteni, 1995; hereafter CM95). 

Most of these simulations (CM93, CM95, MLC94, SM94) were carried 
out using the Smoothed Particle Hydrodynamics (SPH)
technique (Monaghan 1985, 1992). It is found to be a very powerful tool for 
astrophysical applications, particularly when proper care is taken to
ensure the energy and angular momentum conservation. This was done
by implementing the code in cylindrical coordinates where each particle is
assumed to be an axisymmetric torus (Molteni \& Sponholz, 1993). The results
obtained by these simulations generally 
agree with the analytical works, in terms of
the shock location and strengths, etc. However, the SPH methods often are
charged of being unduly dissipative and it is essential to quantify the 
behavior of SPH
by comparing with analytical work, as well as other finite different methods.
Recently, Ryu \etal (1995) have considered 
a problem related to accretion of sub-Keplerian inflow
very similar to MLC94 using a completely different code. 
This code is an Eulerian finite difference code based on the 
Total Variation Diminishing (TVD) scheme developed
by Harten (1983). The code is found to conserve energy and angular
momentum sufficiently accurately, but it has not been tested yet against
analytical works in the non-linear solutions of accretion processes 
of quasi-spherical flows. 
In the present paper, we shall compare the results 
from the TVD code as well as the SPH code with the 
analytical solution of accretion flows in the black hole geometry. 

A few other comparisons between SPH and finite difference codes
have already been made in the literature. 
Steinmetz \& M\"uller (1993) compared one-dimensional
plane-parallel hydrodynamic shock solutions and spherical cloud collapse
and suggested some improvements in the SPH technique to minimize dissipation.
Davies \etal (1993) compared collisions of a main sequence star
with a white dwarf and showed a satisfactory performance of SPH as well
as the finite difference code.
Laguna, Miller \& Zurek (1993) compared a general relativistic SPH 
code for one-dimensional plane-parallel shock solutions and 
spherical inflows (with and without
pressure) with an analytical model and found a very good agreement. 
Kang \etal (1994) compared
six different codes among each other including the TVD code and
two versions of SPH applied to cosmological hydrodynamics,
but did not directly compare numerical results 
to any analytical solution. However,
comparisons of codes and analytical work in relation to {\it rotating}
inflows have not been performed yet. We set out to do this here.

The plan of the present paper is the following. We very briefly
present the general procedure of obtaining analytical solutions
(with or without shocks)
in the next Section. In \S 3, we present model equations 
used in the time dependent simulations and describe the numerical schemes
briefly.  In \S 4, we present one-dimensional solutions
of both codes and discuss merits and demerits. In \S 5,
we present comparisons of simulations in two-dimensions. Finally, in \S 6,
we make concluding remarks.

\section {MODEL EQUATIONS FOR ANALYTICAL STUDY}

We assume a thin, rotating, adiabatic accretion or wind near a black hole.
We take the Newtonian model for the non-rotating central compact object 
as given in terms of the Paczy\'{n}ski \& Wiita (1980)
potential. We also assume a polytropic equation of state for the
accreting (or, outflowing) matter, $P=K \rho^{\gamma}$, where,
$P$ and $\rho$ are the isotropic pressure and the matter density
respectively, $\gamma$ is the adiabatic index (assumed in this
paper to be constant throughout the flow, and is related to the
polytropic index $n$ by $\gamma = 1 + 1/n$) and $K$ is related
to the specific entropy of the flow $s$.
We assume that entropy, and thus $K$ can vary
along a flow line only at a shock. We ignore dissipation, so the specific 
angular momentum $\lambda\equiv xv_{\theta}$ is also constant everywhere.
A complete solution of the
stationary model requires the equations of energy, momentum
and mass conservation supplemented by the transonic conditions at
the critical points and the Rankine-Hugoniot conditions at the
shock. The general procedure followed is the same as is presented
in Chakrabarti (1989). In this Section, we present equations
which are the same as in Chakrabarti \& Molteni (1993) where vertical thickness
is chosen to be constant.
The last assumption allows one to test the TVD and SPH codes in one-dimension. 

We use the mass of the black hole $M$, the velocity of 
light $c$, and the Schwarzschild radius $x_g=2GM/c^2$ as the units of 
mass, velocity and distance respectively. The dimensionless energy 
conservation law can be written as
$$
{\cal E} = \frac{v_x^2}{2}+\frac{a^2}{\gamma-1}
+ \frac{\lambda ^2}{2x^2}+g(x).
\eqno{(1)}
$$
Here $g(x)$ is the radial force potential, which in the pseudo-Newtonian
model takes the form: $g(x) = - 1/2(x-1)^{-1} $.
Here, $v_x$ and $a$ are the non-dimensional radial and the sound
velocities, $x$ is non-dimensional radial distance. Apart from an unimportant
geometric factor, the mass conservation equation is given by
$$
{\dot M} = v_x \rho x h_0,
\eqno{(2)}
$$
where $h_0$ is the constant half-thickness of the flow. 
The shock conditions which
we employ here are the following (subscripts `$-$' and `$+$' refer to
quantities before and after the shock): The energy conservation equation
$$
{\cal E}_+ = {\cal E_-},
\eqno{(3a)}
$$
the pressure balance condition
$$
P_+ + \rho_+ v_{x+}^2 = P_- + \rho_- v_{x-}^2
\eqno{(3b)}
$$
and the mass conservation equation
$$
\dot M_+ = \dot M_-.
\eqno{(3c)}
$$
In order to have a shock, the flow must be supersonic,
i.e., the stationary flow must pass through a sonic
point. The sonic point conditions are derived following the
usual general procedure (C89) and one obtains the sonic
point conditions as
$$
v_c(x_c)=  a_c(x_c),
\eqno{(4a)}
$$
and
$$
a_c^2(x_c)={\frac{\lambda_K^2-\lambda^2}{x_c^2}}_.
\eqno{(4b)}
$$
The subscript $c$ denotes quantities at the critical points.
Here, $\lambda_K (x) = (x/2)^{1/2}/(1-1/x)$ is the Keplerian
angular momentum.
Eqs. (1), (2), (3a-c), (4a-b) are solved simultaneously 
to obtain the full set of solutions which may include shocks.

Whereas above equations are valid for a steady state flow
with constant height, they are easily generalized for a flow in vertical
equilibrium. In this case, the flow height $h_0$ of Eq. (2) 
is replaced by the 
flow thickness $H(x)\sim a x^{1/2} (x-1)$ appropriate for a disk
in vertical equilibrium and the shock condition is imposed over the
integrated pressure and density. Details are in C89 
and we do not discuss them here.

It is not necessary that a flow contains shocks at all. Usually
after the flow deviates from a Keplerian disk, it may pass through the
inner sonic point straight away into the black hole, 
or, through the outer sonic point or both (when
shocks join these two types of flows). The actual fate of the flow
depends on two parameters, the specific angular momentum and the
specific energy (C89). In fact, even the 
outer sonic point may not exist for flows with a large polytropic
index (e.g. $\gamma=5/3$) unless the flow is very thin (MSC96).
Most general and complete global solution topologies of viscous transonic flows 
(Chakrabarti 1996, hereafter C96, and references therein) show
more complex behavior since the presence or absence of transonic
solution depends on efficiencies of cooling and heating processes
and well as the viscosity.

\section{EQUATIONS AND SCHEMES USED IN THE NUMERICAL STUDY}

The TVD code used in this paper is exactly same as that used and described
in details in Ryu \etal (1995), except for the form of the gravitational
force which is modified here according to pseudo-Newtonian potential
(Paczy\'nski \& Wiita, 1980).
The scheme was originally developed by Harten (1983).
It is an explicit, second order accurate scheme which
is designed to solve a hyperbolic system of the conservation equations,
like the system of the hydrodynamic conservation equations.
It is a nonlinear scheme obtained by first modifying the flux function
and then applying a non-oscillatory first order accurate scheme to
get a resulting second order accuracy.
Thus the key merit of this scheme is to achieve the high resolution of a
second order accuracy, while preserving the robustness of a non-oscillatory
first order scheme.

The equations solved numerically with the TVD code are written in 
vector form using non-dimensional units as:
$${\partial{\bmit q}\over\partial t}+{1\over x}{\partial\left(x
{\bmit F}_1\right)\over\partial x}+{\partial{\bmit F}_2\over\partial x}
+{\partial{\bmit G}\over\partial z} = {\bmit S},
\eqno(5a)$$
where the state vector is
$${\bmit q} = \left(\matrix{\rho\cr
                          \rho v_x\cr
                          \rho v_{\theta}\cr
                          \rho v_z\cr
                          E\cr}\right)_, \eqno(5b)$$
the flux functions are
$${\bmit F}_1 = \left(\matrix{\rho v_x\cr
                         \rho v_x^2\cr
                         \rho v_{\theta}v_x\cr
                         \rho v_z v_x\cr
                         (E+p)v_x\cr}\right)\qquad
{\bmit F}_2 = \left(\matrix{0\cr
                          p\cr
                          0\cr
                          0\cr
                          0\cr}\right)\qquad
{\bmit G} =  \left(\matrix{\rho v_z\cr
                         \rho v_x v_z\cr
                         \rho v_{\theta} v_z\cr
                         \rho v_z^2+p\cr
                         (E+p)v_z\cr}\right)_, \eqno(5c)$$
and the source function is
$$
{\textfont1 = \twelvei
      \scriptfont1 = \twelvei \scriptscriptfont1 = \teni
       \def\mit{\fam1 }
{\bmit S} =  \left(\matrix{0\cr
                ~~~\cr
                {\rho v_{\theta}^2\over x}
                -{\rho x\over2\left(\sqrt{x^2+z^2}-1\right)^2\sqrt{x^2+z^2}}\cr
                ~~~\cr
                ~~~\cr
                -{\rho v_x v_{\theta}\over x}\cr
                ~~~\cr
                -{\rho z\over2\left(\sqrt{x^2+z^2}-1\right)^2\sqrt{x^2+z^2}}\cr
                ~~~\cr
                ~~~\cr
                -{\rho \left(xv_x+zv_z\right)\over
                2\left(\sqrt{x^2+z^2}-1\right)^2\sqrt{x^2+z^2}}\cr}\right)_.}
\eqno(5d)$$
Here, energy density $E$ (without potential energy) is defined as,
$E=p/(\gamma-1)+\rho(v_x^2+v_{\theta}^2+v_z^2)/2$.

Note that, in the case of axisymmetric flow without viscosity,
the azimuthal momentum equation is simply the conservation of angular momentum
$${d\lambda\over dt}=0_. \eqno(6)$$
So for the problem considered in this paper, the equation for the azimuthal
momentum can be decoupled from the rest.
But the TVD code used in this paper was designed to solve the whole set of
the equations including that for the azimuthal momentum, so as to be able to
handle the general cases (such as viscous flows)
where the angular momentum could be transported from one region to 
another.  As a result, the TVD calculations do not conserve the angular momentum
exactly but have an error typically less than a couple of percent in
the angular momentum conservation.

The equations governing the SPH simulations are already provided in
CM93, and MLC95, and we do not repeat them here.

In what follows, we present 
the results of the calculations using the SPH and TVD codes comparing them 
with the analytical work in one-dimensional cases and with each other in
two-dimensional cases.

\section{COMPARISON OF NUMERICAL SIMULATION RESULTS IN ONE-DIMENSION}

Physically, a shock can form when the rotating flow is close to the centrifugal
barrier. However, because the flow can have a significant pressure,
it is not essential that the angular momentum should
be so high as to hit an actual barrier. That is, the flow angular momentum
could be much lower than the marginally stable value ($\sim 1.83$ in our units).
In the current simulation, we choose $\lambda=1.8$, slightly below the 
marginally stable value. In the following comparison, we present the 
TVD calculation with $512$ grid points
between $0$ to $x=50$ (Fig. 1 and Fig. 2b) or $0$ to $x=90$ (Fig. 2b), 
and the SPH calculation with $\sim 560$ particles (of size $h=0.3$)
in Fig. 1 and Fig. 2a, and $\sim 1080$ particles (of size $h=0.25$)
in Fig. 2b.
Matter is injected at the outer boundary located at $x=50$ or $90$ subsonically
with the velocity and sound speed obtained from the analytical
solution. At the inner boundary, an absorption boundary condition is used
to mimic the black hole horizon.
In the TVD simulations the inner edge is kept at $x=1.5$, and
in the SPH simulations it is kept at $x=1.25$. (Horizon is at $x=1$).
We choose the adiabatic index $\gamma=4/3$, which is 
appropriate for very relativistic flows or very optically thick radiation 
pressure dominated flows. This index is also valid for optically thin flows
(perhaps in the presence of magnetic fields) with some cooling effects
(such as the Comptonization of external soft photons).
The goal was to see to what extent shocks are captured by the flow.
In the TVD simulations we have tested the one dimensional
code with 128, 256, 512 and 1024 uniformly spaced grid points 
whereas in the SPH simulations we have tested the code
with various particle sizes $h$.
We find the result of varying resolutions to agree with each other
very well, though the thickness of the shock could vary with the
grid resolution. Thus, we are convinced that the code is definitely
converged at the grid resolution presented.

Analytical solution of a given problem requires two conserved
quantities, such as the specific energy ${\cal E}$ and 
angular momentum ${\lambda}$. The entire solution, 
including the locations of sonic points and the shock,
is determined by these two parameters (C89). If the flow is
viscous, the viscosity parameter is also required (CM95, C96 and 
references therein) and in return, the location where the flow 
joins with a Keplerian disk comes out as an eigenvalue.
In the first test case, we choose the energy parameter
${\cal E}=0.0363$ with angular momentum $\lambda=1.8$ and the
outer boundary at $x=50$. 
In Fig. 1, we superimpose the numerical simulation results 
on the analytical result. For clarity near the inner egde
we only show the region $x\leq35$  (for $x>35$ agreement
is perfect). In this case, the flow starts out subsonically,
presumably from a Keplerian disk,
then enters through the outer sonic point,
passes through the shock and then subsequently passes through the
inner sonic point before entering the black hole. The upper panel shows
the variation of density and the lower panel shows the
variation of the Mach number with the 
radial distance ($x$). The solid curve
is the analytical solution, whereas the long dashed curve and the short
dashed curves are the results of the TVD calculation
and the SPH calculation respectively. The agreement is clearly very good.
The shock is thicker in the SPH calculation than in the TVD one.
Also the density peak around $x\sim3.5$ is slightly
underestimated in the SPH calculation.

In the analytical solution,
the locations of the outer sonic point $x_{o}$, the inner sonic point
$x_{i}$ and the stable shock $x_{s3}$ (using the notations of
C89, CM93), for the flow parameters mentioned above,
are at $27.9$, $2.563$ and $7.8896$ respectively.
In the TVD simulation, these numbers were $27.97$, $2.57$ and $7.98$
respectively, while in the SPH simulation we find these numbers to be
$26.78$, $2.571$ and $8.46$ respectively. The exact shock
location is probably in error since the shock is somewhat diffuse,
more so in the SPH simulation.

It is known that for every angular momentum (within a range) at the
outer boundary, there is a range of specific energy  of the flow,
for which the shock conditions are satisfied in accretion flows
(see, Fig. 4 of C89). If the energy is smaller 
than the lower limit of the range, the flow only passes through the
outer sonic point and no shocks are present.  
Similarly, if the energy is higher than
the upper limit of the range, the flow would only pass through the
inner sonic point and the question of shock formation does not appear.
This is true even if the flow of outside the shock regime
has two saddle type sonic points.
We present now two such examples where the flow can pass only through
one sonic point at a time.
In Fig. 2a, we present a comparison of solutions which has only
the inner sonic point at $x_i=2.46$. The energy parameter in this
case is ${\cal E}=0.07$ with $\lambda=1.8$. The outer boundary  for the simulation is 
chosen at $x=50$.
In Fig. 2b, we present the comparison of solutions with the outer sonic point
located at $x_o=77.615$. The energy parameter in this case is
${\cal E}=0.015$, with $\lambda=1.8$. The outer boundary for the
simulation is chosen at $x=90$.
Again the plots show only the region with $x\leq35$ for clarity
of comparison close to the black hole.
The solid, long-dashed and short dashed curves are the analytical,
TVD and SPH solutions respectively.
The TVD results agrees very well with the analytic solutions.
The agreement of the SPH results is clearly acceptable, although not quite as good
as that of the TVD ones.

\section{COMPARISON OF SHOCK STUDIES IN TWO-DIMENSIONAL FLOWS}

A realistic accretion disk is expected to be three-dimensional, of course.
Assuming axisymmetry, the problem is reduced to two-dimensions.
It is difficult to solve analytically the problem in full generality
even in two-dimensions. The nearest case that is solved analytically is
the inflow which is in instantaneously in vertical equilibrium (C89, C96).
A comparison of such an analytical solution with 
the two-dimensional SPH solution (MLC94)
suggests that the flow displays more complex behavior, because of
the presence of turbulence in the  backflow which is
produced as the flow hits the centrifugal barrier. The turbulent pressure
seems to be comparable to other pressure effects, such as thermal 
and ram pressure, and it changes the location of the shock significantly.
It also increases the chance of shock formation. The parameter space
within which steady shocks may develop appeared to be
much larger in presence of turbulence (MLC94).

For the simulations of the two-dimensional quasi-spherical flow, 
we choose to place the outer boundary at $x=50$. The height
of the disk is $56$ at the outer boundary (assuming vertical equilibrium).
The specific angular momentum
of the flow is chosen to be $\lambda=1.65$. The radial (in the spherical sense)
velocity components is chosen to be constant at all height $v_r=
(v_x^2+v_z^2)^{1/2}=0.068212$.
We also choose the adiabatic index $\gamma=4/3$ as before.
On the equatorial plane, the energy parameter chosen is ${\cal E}=0.004$.
This implies a sound speed of $a=0.061463$ on the equatorial plane.
We employ an isothermal outer boundary condition (namely, the same sound speed
at all height). The matter is injected at the outer boundary in both the codes
and is absorbed at the inner edge in the same way as in the one-dimensional case.

With the specific energy and angular momentum
chosen as above (${\cal E}=0.004$, $\lambda=1.65$), according
to the model flow in vertical equilibrium (C89)
there is no analytical shock solution in the {\it equatorial
plane} in the absence of turbulence pressure. However, 
the specific energy increases with vertical height
(with the decrease in potential energy)
and the flow comes within the regime of shock formation even without
the turbulence (Fig. 8c of C89). Thus a shock is naturally expected at a higher
elevation. Since the gravitational force decreases with vertical height,
it requires less centrifugal force to form a shock. Hence the
shock is expected to be at a larger distance $x$ at higher elevations. In other 
words, it is expected that the shock would be bent outward as
one deviates away from the equatorial plane. 

There are two important features of a rotating, thick accretion flow:
one is the funnel wall and the other is the centrifugal barrier.
Roughly speaking, the funnel wall is described by the surface
($x_f, \ z_f$) which is the surface of vanishing effective potential
$$
\Phi_{eff} = -\frac{1}{2(r_f-1)} + \frac {\lambda^2}{2x_f^2} = 0,
\eqno{(7)}
$$
where, $r=(x^2+z^2)^{1/2}$ is the spherical distance.
On the other hand, the centrifugal barrier 
($x_{cf}, \ z_{cf}$) is governed by the competition between the
centrifugal force and gravitational force:
$$
-\frac{1}{2(r_{cf}-1)^2} + \frac {\lambda^2}{x_{cf}^3} = 0.
\eqno{(8)}
$$
Matter stays bound up to the effective potential barrier, or the
so-called funnel wall. However, the centrifugal 
force dominates over gravity between
the funnel wall and the centrifugal barrier. The accretion flow
first hits the effective potential barrier 
and piles up behind it. A part of the flow
moves backward, and its interaction with the infalling matter 
causes turbulence. Combined effects of thermal pressure $P$
(whose presence is mainly due to centrifugal force anyway) and turbulent
pressure $P_t$ `brakes' the incoming flow  which then undergoes
a shock transition.
If the piled up `atmosphere' of the black hole thickens with time, either
due to a very strong centrifugal barrier or because of faster viscous
transport of angular momentum in this region, the shock propagates
outwards and is not steady (CM95). But in presence of smaller viscosity
the shock can be steady and forms another surface where the pressure
balance condition (Eq. 3b) is satisfied in the thick flow:
$$
P_+(x,z) + P_{t+}+ \rho_+(x,z) v_+(x,z)^2 =
P_-(x,z) + P_{t-}+ \rho_-(x,z) v_-(x,z)^2.
\eqno{(9)}
$$
As mentioned earlier, the shock is expected to bend outwards,
roughly following the contour of the centrifugal barrier (Eq. 8).

Below, we show that our simulations using two codes
capture all the three surfaces described by Eqs. 7-9 very unambiguously.

The TVD calculation for the two-dimensional flow has been done with
$128\times256$ cells in a box of $50\times100$.
The SPH calculation has been done with $\sim 12800$ particles in the same box.
The results show that, after the initial set-up period with
$t\lsim5\times10^3$, the matters settles on a steady-state solution.

Figs. 3(a-b) plot the density contours of the TVD (Fig. 3a) 
and SPH (Fig. 3b) calculations in a box of $50\times50$ at $t=2\times10^4$. 
Both the solutions reached steady states (in large scale behavior)
much before this time. At the outer boundary, density is chosen 
to be unity in both the codes.  Within the box of $50 \times 50$,
the number of TVD grids are $128\times128= 16384$, out of which 
a couple of thousands are basically empty (within the funnel 
wall). In SPH calculation, on the other hand, within $50 \times 50$ box there
were $11229$ particles, all participating in the dynamics of the flow.
Thus, the resolutions in the two calculations are probably comparable,
though SPH resolution is higher closer to the black hole due to
accumulation of more particles there.
In both the figures, the contour level of the minimum density 
is $0.0261$.  Successive contours have a density ratio of $1.2$.
Only some of the contours are marked for clarity. 
The plots of both the simulations have the same contour levels.

There are clearly general agreements
between the two calculations. Both the simulations show clear presence of 
the funnel wall. (Ruggedness of the contours in Fig. 3b is an artifact
of interpotation of SPH results on equal spaced grids required for
contour plotting.) The most visible structure is the quasi
parabolic stationary accretion shock which runs vertically 
and bends outwards as discussed above.
In the SPH simulation, the angular momentum of each particle is preserved
by construction (as each particle is described by an axisymmetric torus),
but in the TVD code it can be redistributed due to numerical diffusion.
A closer examination of the results has shown that the TVD code preserves
angular momentum to within a couple of percent. In fact, it seems
to have a little less average angular momentum near the shock. This may make the
centrifugal force somewhat weaker and the paraboloidal shock may be
located inwards because of this. Such a dependence of the shock location with 
angular momentum also conforms with analytical understanding (e.g., 
Fig. 8c of C89). Note that due to general reduction of angular
momentum, matter tended to pile up more closer to the black hole
in TVD simulation. This is evidenced by a comparison
of the locations of a given density contour on both the figures.

Both simulations show the presence of another oblate spheroidal
shock near the equatorial plane. Both touch the equatorial plane $(z=0)$
at roughly the same location, at $x\sim24$ in the TVD simulation
and at $x\sim26$ at the SPH simulation. This shock near the equatorial plane
is a result of the combination of backflow and pressure effects.
Whereas in TVD code the two shocks (the parabolic and the spheroidal
shocks) cross each other possibly because the
paraboloidal shock may have moved inwards, in the SPH results they just
touch on the equatorial plane.  Away from the equatorial plane, this
shock is formed by the accelerated post paraboloidal shock flow
coming from the higher elevation.
In general, the shocks are better resolved in TVD than in SPH.

Figs. 4 plot the velocity fields of the TVD calculation (Fig. 4a)
and the SPH calculation (Fig. 4b).
In the TVD plot, the cells with the density higher than
$0.0261$ of the inflow density at the outer boundary have been
shown with arrows placed on alternate grid points for clarity, 
while in the SPH plot all densities are used. But arrows have been
put on every fifth particle (of the final sequence) for clarity.
We also superimpose two surfaces, the funnel wall
described by Eq. 7 (inner curve) and the centrifugal
barrier by Eq. 8 (outer curve) on these velocity fields. 
In the TVD calculations some inflow material close to the funnel wall
is mixed with the background material of zero angular momentum, resulting
in material with smaller angular momentum than the inflow material.
This is a disadvantage of grid based finite difference codes.
On the other hand, in the SPH code the inflow material is represented by
particles and there is no need for any background material.
As a result, in the SPH plot all material is outside the funnel wall,
while in the TVD plot some matter penetrates inside the funnel wall.
The presence of the
prominent centrifugal barrier in the SPH plot is probably
due to the strict conservation of angular momentum in SPH.
However, since the shock is better resolved in TVD, the velocity 
vectors reflect better the shock transition of the parabolic
accretion shock (from supersonic long vectors to subsonic short vectors).
The back flow toward the positive $x$-direction inside the vortex is
clearly visible in the TVD calculation, but does not exist in the SPH 
calculation.  The clumsiness of the
arrows in SPH closer to the hole is because too many particles are
present there. But the flow is perfectly smooth and well behaved.

The test calculation with a lower resolution using $64\times128$ cells using 
a TVD code has shown the same structures as those in Figs. 3a and 4a with
the positions of shocks and vortex agreeing reasonably well. SPH simulation
has also been carried out with larger particle size $h\sim 1$ and produce
thicker shock wave, but at the same location. 
Although this is partly because our calculation has produced relatively
simple structures, we think that the two-dimensional calculation presented
in this paper show converged results.

\section{CONCLUDING REMARKS}

In the present paper, we have compared for the first time 
analytical results of inviscid transonic flows with the results of two
hydrodynamic codes based on completely different mathematical
techniques. We have found that, roughly
both the codes agree with each other and with analytical solutions
quite well, though the details vary. 
In one-dimensional calculations, both the codes
agree very well with the analytical solution of density and
Mach number distributions. They reproduce the locations
of the sonic points and the shocks with high accuracy.
For the first time, we have also been able to actually follow 
subsonic inflow at the outer boundary while it crosses the outer sonic point 
in its inward journey with (Fig. 1) or without shocks (Fig. 2). 
In two-dimensional calculations, both the
codes have shown the formation of the funnel wall and the centrifugal
and turbulent pressure supported shocks.
In TVD calculation angular momentum
is conserved within a couple of percent, which might have caused
the accretion shock to form slightly inside. The SPH 
technique conserves angular momentum strictly. On
the other hand, in the TVD 
calculation the shocks are well resolved while in the SPH
calculation the shocks are wider. 
In two-dimensions, a complete analytical theory does not exist.
However, extrapolated understanding of one-and-a-half dimensional
flow behavior (C89), such as the curved vertical shock, funnel
wall, and the centrifugal barrier, agrees well with what 
we see in our simulations. Based upon this experience, we believe
that just as the comparison of plane-parallel shock results
constitutes a definitive test for accuracy of 
a code written in the Cartesian coordinate, any code written in spherical
or cylindrical co-ordinate should be tested against our 
shock solutions to measure the accuracy of such codes.

A realistic two-dimensional flow at the outer boundary located at a 
large distance may either be sub-Keplerian
(as probably in active galaxies) or may be Keplerian (in binary systems
and active galaxies). But in both the cases the inflow at the outer
boundary is expected to be subsonic. We have already presented
examples of stable one-dimensional solutions, with or without shocks
in our simulations, which are subsonic to begin with.
Fully two-dimensional simulations with the subsonic
outer boundary condition (which necessarily means that the boundary 
be located much farther out beyond the outer sonic point) 
require more efforts and the computational time.
Similarly, our present simulations were done for parameter
ranges producing steady solutions. In reality, some solutions
may be inherently time dependent showing oscillations of
various time scale (as in MSC96, Ryu \etal 1995). 
Such simulations are presently in progress and 
will be described in a subsequent paper.

\acknowledgments 

The work by DR was supported in part by the Basic Science Research Institute
Program, Korean Ministry of Education 1995, Project No.~BSRI-95-5408.
The authors thank the referee for suggesting improvements in overall
presentation of the paper.

\clearpage

\begin{center}
{\bf FIGURE CAPTIONS}
\end{center}
\begin{description}

\item[Fig.~1] Comparison of analytical and numerical results
in a one-dimensional accretion flow which allows a standing shock.
${\cal E}=0.036$ and $\lambda=1.80$ are used. The long
and short dashed curves are the results of the TVD and SPH simulations
respectively. The solid curve is the analytical result for the
same parameters. Upper panel is the mass density 
in arbitrary units and the lower panel is the Mach number of the flow.
Here the flow passes through the outer sonic point (at $x_o=27.9$), 
then through a shock (at $7.89$), and finally through the
inner sonic point (at $2.563$). 

\item[Fig.~2a-b] As in Fig. 1, but for flows without shocks while keeping
$\lambda=1.80$.
In (a), the flow has ${\cal E}=0.07$ and it passes through only the
inner sonic point at $x_i=2.46$. In (b), 
the flow has ${\cal E}=0.015$ and it
passes through only the outer sonic point at $x_o=77.615$.

\item[Fig.~3a-b] Density contours of (a) the TVD simulation and
(b) the SPH simulation in the $x-z$ plane at $T=2 \times 10^4$.
Same set of contour levels are used. In both the
cases the big paraboloidal accretion shock dominates the flow. Also
seen are the oblate spheroidal shock near the equatorial plane which 
forms mainly due to turbulence pressure of the backflow. The 
parameters used in the calculations are ${\cal E}=0.004$ 
and $\lambda=1.65$ in the outer boundary on the equatorial plane.

\item[Fig.~4a-b] As in Fig. 3 but the velocity fields are shown. 
Super-imposed by solid curves are the locations of the
funnel wall (inner curve) and the centrifugal barrier (outer curve)
respectively. 

\end{description}


\clearpage

\begin{thebibliography}{}

\bibitem[ ]{ } Chakrabarti, S. K., 1989, ApJ, 347, 365  (C89).

\bibitem[ ]{ } Chakrabarti, S. K., 1996, ApJ, (June 20th issue, in press).

\bibitem[ ]{ } Chakrabarti, S. K., \& Molteni, D., 1993, ApJ, 417, 671 (CM93).

\bibitem[ ]{ } Chakrabarti, S. K., \& Molteni, D., 1995, MNRAS, 272, 80 (CM95).

\bibitem[ ]{ } Chakrabarti, S. K., \& Titarchuk, L. G., 1995, ApJ, 455, 623.

\bibitem[ ]{ } Davies, M. B., Ruffert, M., Benz, W., \& M\"uller, E.,
1993, A \& A, 272, 430.

\bibitem[ ]{ } Harten, A., 1983, J. Comput. Phys., 49, 357.

\bibitem[ ]{ } Hawley, J. F., Smarr, L. L., \& Wilson, J. R., 
1984, ApJ, 277, 296.

\bibitem[ ]{ } Hawley, J. F., Smarr, L. L., \& Wilson, J. R., 
1985, ApJ Suppl., 55, 211.

\bibitem[ ]{ } Kang, H., Cen, R., Ostriker, J. P., \& Ryu, D., 
1994, ApJ, 428, 1.

\bibitem[ ]{ } Monaghan J. J., Comp. Phys. Repts., 1985, 3, 71.

\bibitem[ ]{ } Monaghan J. J., Ann. Rev. Astron. Astrophys., 1992, 30, 543.

\bibitem[ ]{ } Laguna, P., Miller, W.A., Zurek, W.H., \& Davies, M.B.,
1993, ApJ, 410, L83.

\bibitem[ ]{ } Liang, E. P. T., \& Thompson, K. A., 1980, ApJ, 240, 271.

\bibitem[ ]{ } Molteni, D., Chakrabarti, S. K., \& Ryu, D., 1996,
in preparation.

\bibitem[ ]{ } Molteni, D., Lanzafame, G., \& Chakrabarti, S. K., 
1994, ApJ, 425, 161 (MLC94).

\bibitem[ ]{ } Molteni, D. \& Sponholz, H. 1994, in Journal of Italian 
Astronomical Society, Vol. 65-N. 4-1994, Ed. G. Bodo \& J.C. Miller 

\bibitem[ ]{ } Molteni, D., Sponholz, H., \& Chakrabarti, S. K., 
1996, Feb. 1st (MSC96).

\bibitem[ ]{ } Novikov, I., \& Thorne, K. S., 1973, in Black Holes,
eds. C. DeWitt and B. DeWitt (Gordon and Breach, New York).

\bibitem[ ]{ } Paczy\'nski, B., \& Bisnovatyi-Kogan, G. ,
1981, Acta Astron. 31, 1.

\bibitem[ ]{ } Paczy\'nski, B., \& Wiita, P. J., 1980, A\&Ap, 88, 23.

\bibitem[ ]{ } Ryu, D., Brown, G. L., Ostriker, J. P., and Loeb, A., 
1995, ApJ, 452, 364.

\bibitem[ ]{ } Ryu, D., Chakrabarti, S. K., \& Molteni, D. 1996, 
in preparation.

\bibitem[ ]{ } Ryu, D., Ostriker, J. P., Kang, H., and Cen, R., 
1993, ApJ, 414, 1.

\bibitem[ ]{ } Shakura, N. I., \& Sunyaev, R. A., 1973, A\&A, 24, 337.

\bibitem[ ]{ } Sponholz, H., \& Molteni, D., 1994,
MNRAS, 271, 233 (SM94).

\bibitem[ ]{ } Steinmetz, M., \& M\"uller, E., 1993,
A\&Ap, 268, 391.

\bibitem[ ]{} Thompson, J.M.T. and Stewart, H.B., 1985, Nonlinear Dynamics
and Chaos (John Willey \& Sons Ltd.).

\bibitem[ ]{ } Wilson, J. R., 1978, ApJ, 173, 431.

\end{thebibliography}
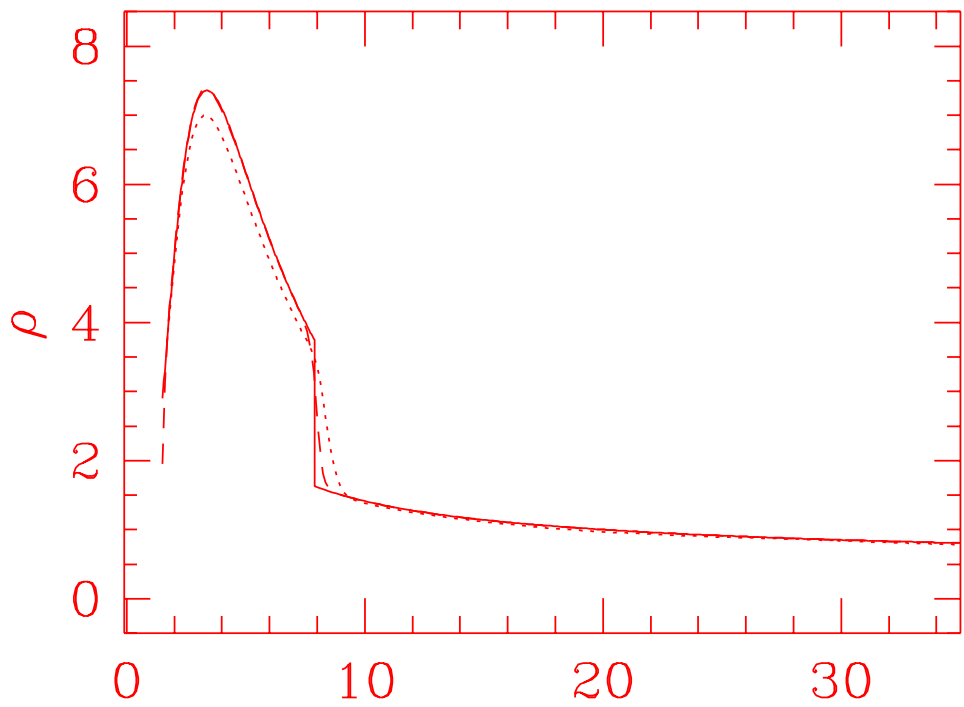
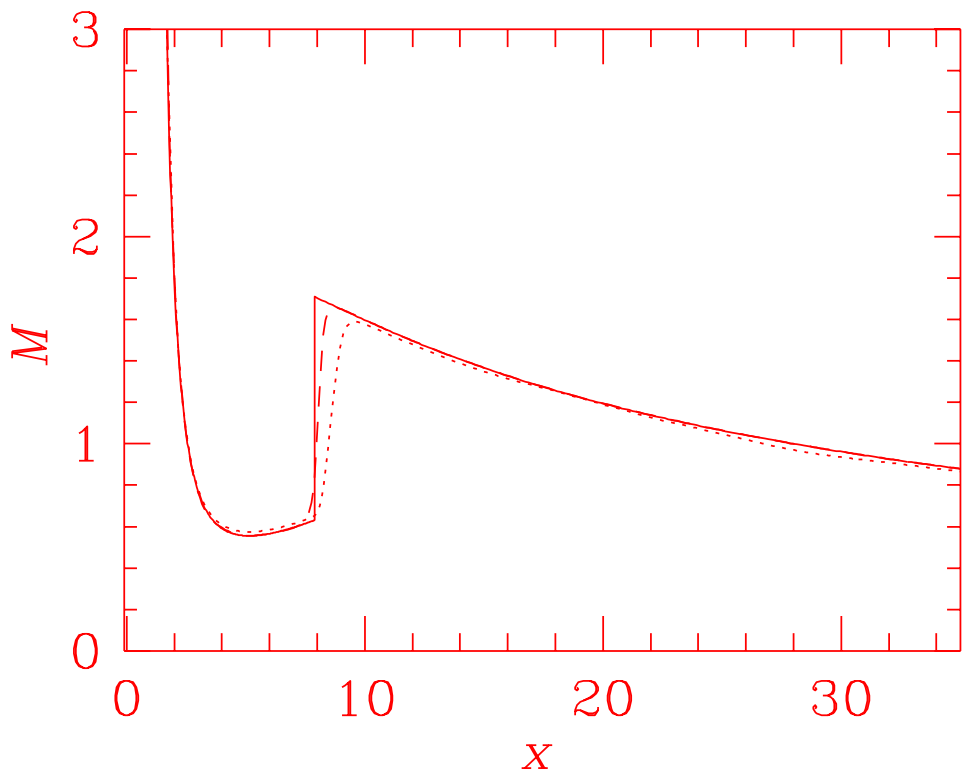
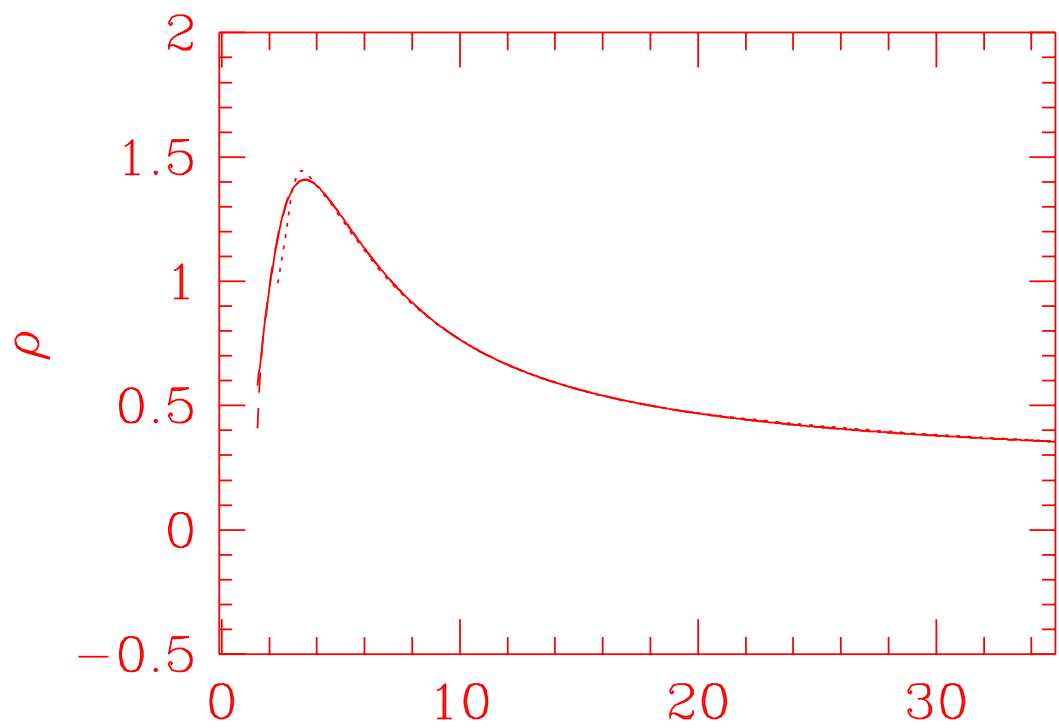
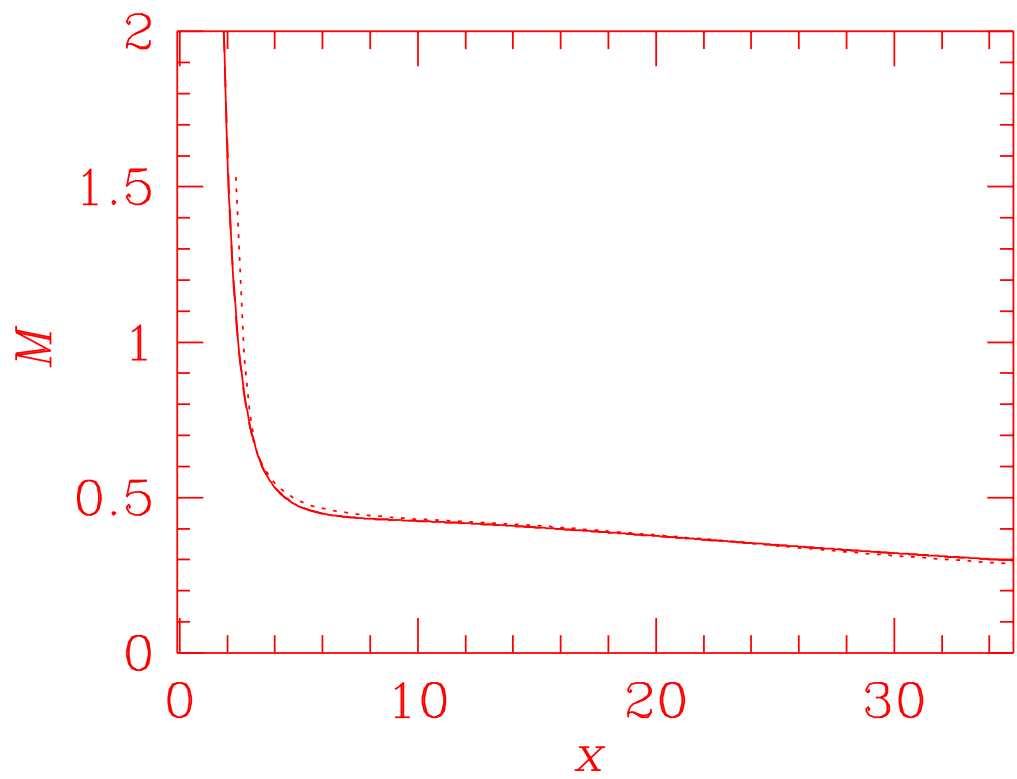
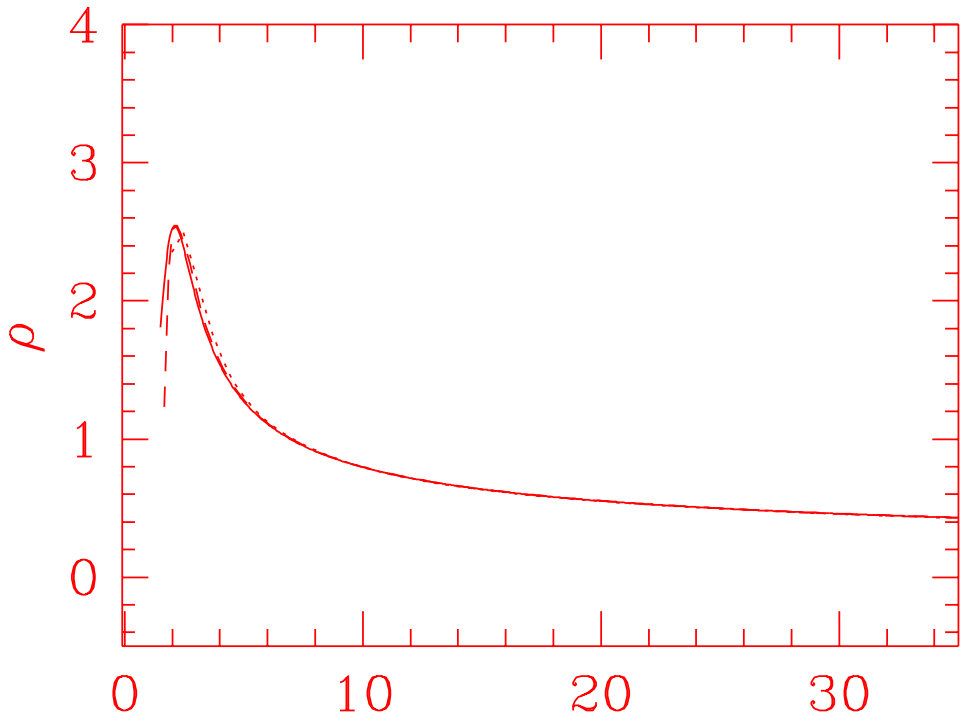
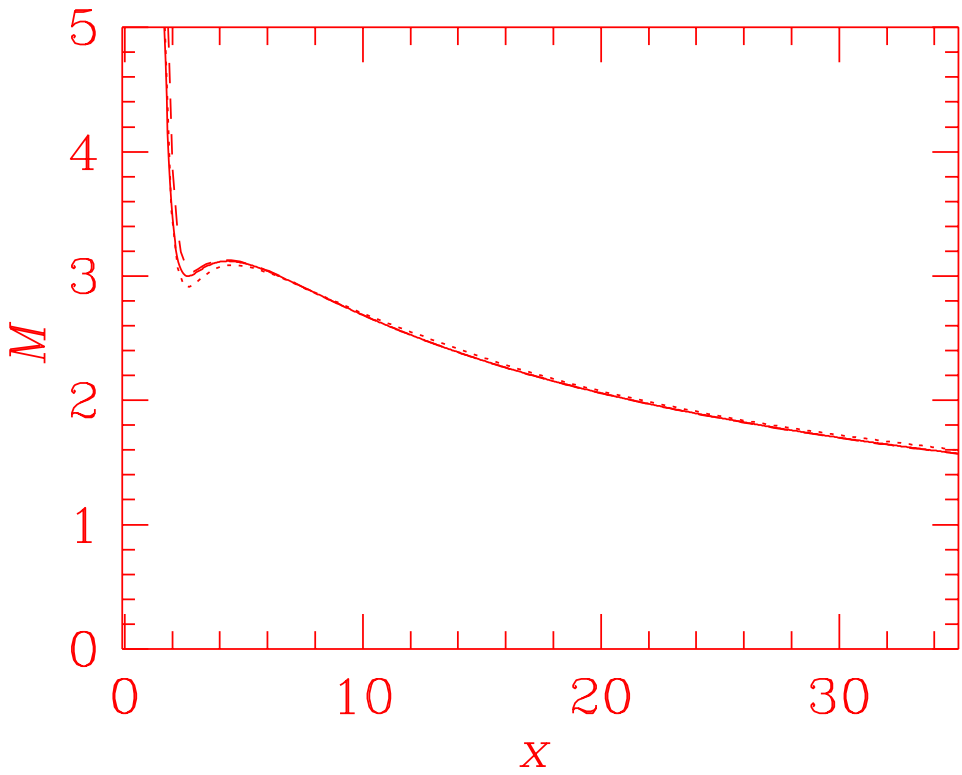
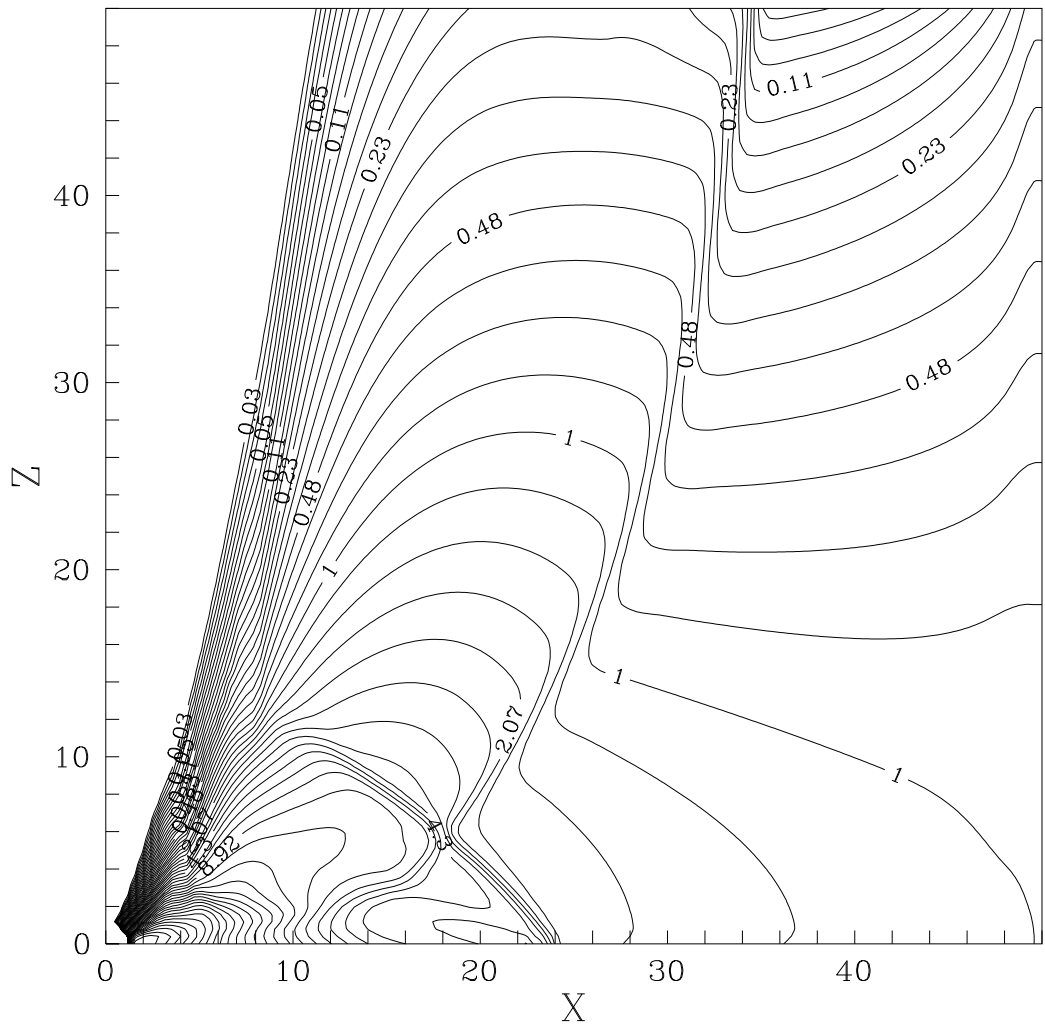
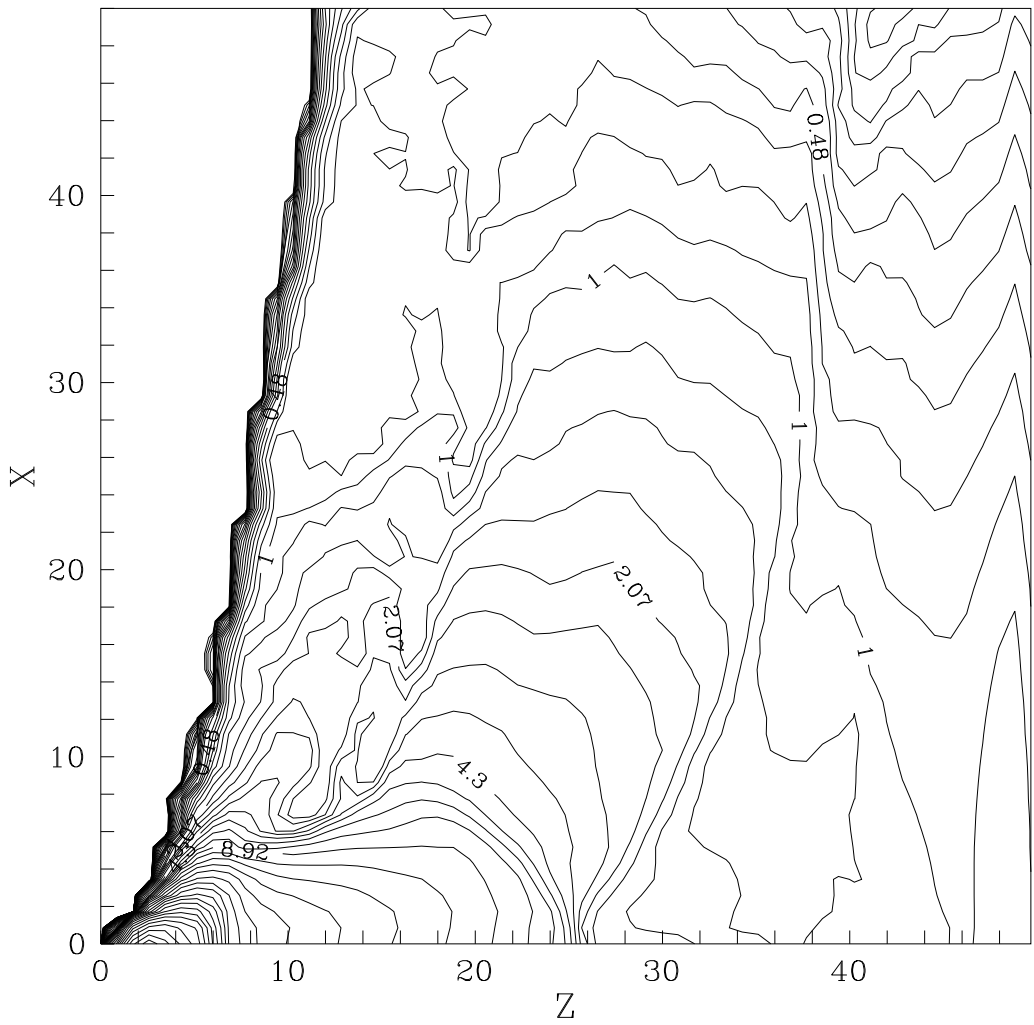
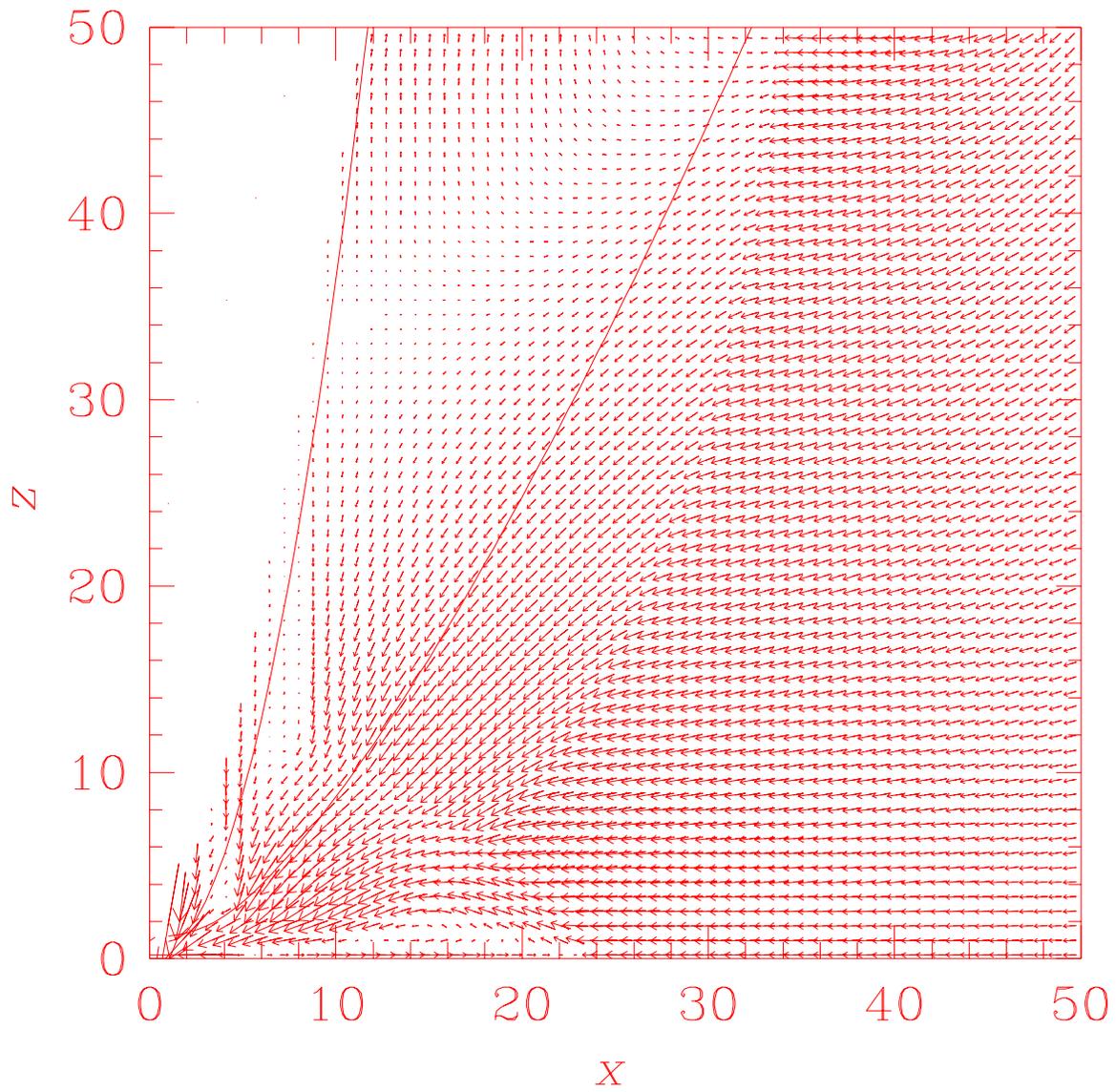
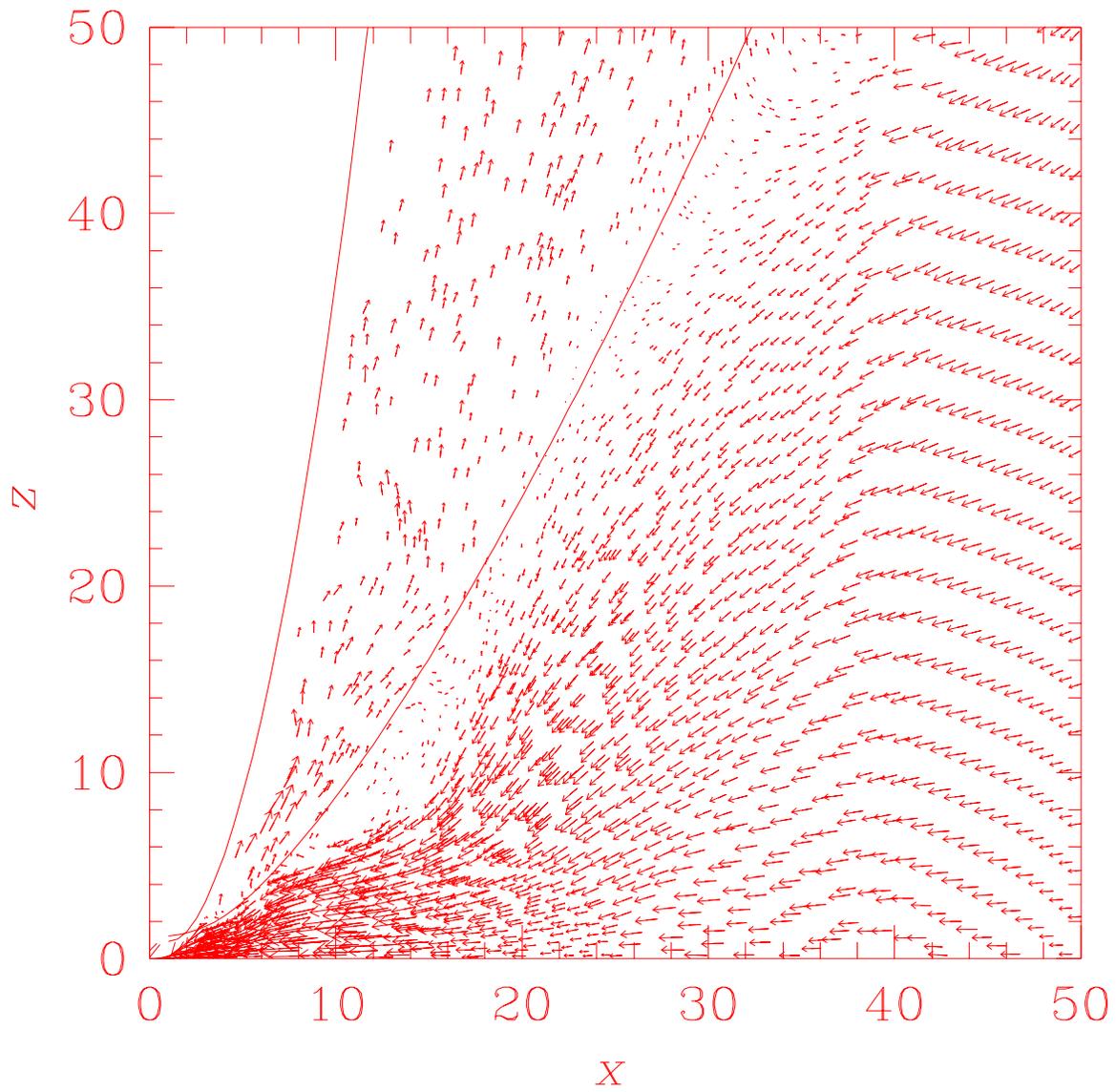
\end{document}